\documentclass{isprs} 
\usepackage{subfigure}
\usepackage{setspace}
\usepackage{geometry} 
\usepackage{epstopdf}
\usepackage[labelsep=period]{caption}  
\usepackage[british]{babel} 
\usepackage[hang]{footmisc}
\usepackage{braket}
\usepackage{amsmath} 
\usepackage{xcolor}
\usepackage{soul}

\begin{document}

\title{Quantum Meets SAR:\\A Novel Range-Doppler Algorithm for Next-Gen Earth Observation}
\date{}

\author{
Khalil Al Salahat\textsuperscript{1}, 
Mohamad El Moussawi\textsuperscript{1}, 
Ali J. Ghandour\textsuperscript{2*}
}

\address{
\textsuperscript{1} MSc students, Geoinformatics Engineering, Politecnico di Milano, Milan, Italy. \\
\textsuperscript{2} National Center for Remote Sensing, CNRS-L, Beirut, Lebanon \\ \textsuperscript{*} Corresponding Author: aghandour@cnrs.edu.lb
}


\icwg{}   

\abstract{
Synthetic Aperture Radar (SAR) plays a vital role in remote sensing due to its ability to capture high-resolution images regardless of weather conditions or daylight. However, to transform the raw SAR signals into interpretable imagery, advanced data processing techniques are essential. A widely used technique for this purpose is the \textbf{Range Doppler Algorithm (RDA)}, which takes advantage of Fast Fourier Transform (FFT) to convert signals into the frequency domain for further processing. However, the computational cost of this approach becomes significant when dealing with large datasets. This paper presents a \textbf{Quantum Range Doppler Algorithm (QRDA)} that utilizes the \textbf{Quantum Fourier Transform (QFT)} to accelerate exponentially by up to $N/(\log N)^2$ processing compared to the classical FFT theoratically. Furthermore, it introduces a quantum implementation of the \textbf{Range Cell Migration Correction (RCMC)} in the Fourier domain, a critical step in the RDA pipeline that realigns the received echoes so that the energy from a target is concentrated in a single range bin across all azimuth positions. The performance of the quantum RCMC is evaluated and compared against its classical counterpart, as an isolated operation, and then as part of the full classical pipeline, demonstrating the potential of quantum computing in advanced SAR imaging. 
}


\maketitle


\section{Introduction}\label{Introduction}
 
\sloppy
\textbf{Synthetic Aperture Radar (SAR)} is a powerful active imaging modality widely used in remote sensing due to its ability to operate independently of weather conditions and ambient lighting. In a typical SAR system, an airborne or spaceborne platform emits microwave pulses toward a target area and records the backscattered echoes. The raw data collected in this way encode information about the reflectivity properties of the scene, and transforming it into a usable image requires sophisticated processing to account for the motion of the radar and the geometry of wave propagation.
The \textbf{Range Doppler Algorithm (RDA)} is one of the most widely used methods for the formation of SAR images. Its computational core relies on the Fast Fourier Transform (FFT) to convert the data into the frequency domain, where key operations such as range compression, \textbf{Range Cell Migration Correction (RCMC)}, and azimuth compression are performed. However, the complexity of FFT $O(n\log n)$, although efficient for moderate datasets, becomes a bottleneck for large-scale or high-resolution SAR applications, where near-real-time processing is often desirable.
In this work, we explore a \textbf{quantum computing} approach to accelerate the RDA by replacing classical FFT-based steps with their quantum counterparts. The \textbf{Quantum Fourier Transform (QFT)} offers a theoretical exponential speedup over the FFT, but this advantage is contingent on maintaining quantum coherence throughout the entire processing pipeline. Measuring quantum states halfway through (e.g., after \textbf{QFT} but before \textbf{RCMC}) collapses the quantum states, negating the speedup by losing all superposition information from before the measurement and limiting it to a singular result. Thus, a practical quantum RDA must integrate all critical steps, including \textbf{RCMC}, within the quantum domain to avoid decoherence and maximize computational gains~\cite{b1}~\cite{b2}.

\begin{figure}[ht!]
\begin{center}
		\includegraphics[width=1.0\columnwidth]{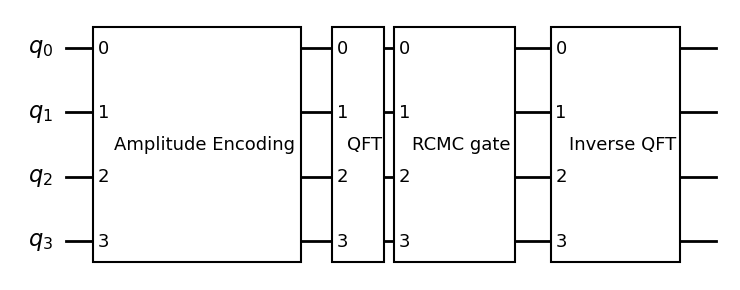}
	\caption{Quantum circuit approach for the Data Encoding and the RCMC gate implementation.}
\label{fig:RCMC}
\end{center}
\end{figure}

To address these challenges, the contribution of this paper is three-fold as follows:
\begin{enumerate}
    \item Efficient quantum encoding, using as few qubits as possible, which allows us to encode N features with only $\log_2(N)$ qubits.
    \item A quantum-domain \textbf{Range Cell Migration Correction (RCMC)} gate, which is a crucial part in the \textbf{Range Doppler algorithm}. Figure~\ref{fig:RCMC} shows the proposed RCMC quantum circuit.
    \item Evaluation of the proposed \textbf{RCMC} gate on real SAR data in two setups: \textit{(i)} isolation  on a 64×64 pixel subset Sentinel-1 SAR data and \textit{(ii)} integration in a full classical RDA.
\end{enumerate}

The rest of this paper is organized as follows: Section~\ref{sec:Fundamentals of Quantum Computing} introduces the basic principles of quantum computing, including qubits, gates, and unitary operators. Section~\ref{sec:Range Doppler Algorithm} reviews the classical Range Doppler Algorithm (RDA). Section~\ref{sec:QFT and QRCMC} presents the proposed quantum circuit implementation of the RDA, with emphasis on the Quantum Fourier Transform (QFT) and the  Quantum Range Cell Migration Correction (QRCMC) gate. Section~\ref{sec:Results} evaluates the performance and correctness of our quantum approach through simulations on real SAR data in isolation and in integration in a full classical RDA. Finally, Section~\ref{sec:Conclusion} summarizes our findings and outlines directions for future research in quantum SAR processing.

\section{Fundamentals of Quantum Computing}\label{sec:Fundamentals of Quantum Computing}


Classical computation is based on bits as its fundamental unit of information, where each bit represents a binary state (0 or 1). For a system with $m$ distinct states, the minimum number of bits required is $n = [\log_2 m]$. In contrast, quantum computation takes advantage of the qubit (quantum bit), which generalizes the classical bit by leveraging superposition and entanglement~\cite{b2}.

\subsection{The Qubit and Superposition}\label{sec:The Qubit and Superposition}
A qubit's state $\ket{\psi}$ is represented as a linear combination of basis states $\ket{0}$ and $\ket{1}$ as shown in Equation~\ref{eq:qubitstates} :

\begin{equation}\label{eq:qubitstates}
\ket{\psi} = \alpha\ket{0} + \beta\ket{1} \text{, where } |\alpha|^2 + |\beta|^2 = 1
\end{equation}

where $|\alpha|^2$ and $|\beta|^2$ tell us the probability of finding $\ket{\psi}$ in the states $\ket{0}$ and $\ket{1}$.
When a qubit is measured, it will only be found to be in the state $\ket{0}$ or the state $\ket{1}$.
This quality of superposition is at the core of \textbf{quantum computing}, allowing us to tap into more possibilities with a single qubit, along with giving us exponential scaling, since $n$ qubits can represent $2^n$ states simultaneously, compared to classical bits, which can only store one state at a time~\cite{b3}.
\subsection{Quantum States and Basis Representations}\label{sec:Quantum States and Basis Representations}

A quantum state $\ket{\psi}$ can be written as a linear combination of a basis set $\ket{v_i}$ with complex coefficients of expansion $c_i$ as shown in Equation~\ref{eq:qubasis}: 

\begin{equation}\label{eq:qubasis}
\ket{\psi} = \sum_{i=1}^{n}c_i\ket{v_i} = c_1\ket{v_1} + c_2\ket{v_2} + ...+ c_n\ket{v_n}
\end{equation}

with $\sum_i |c_i|^2 = 1$. The squared modulus of a given coefficient $c_i$ gives the probability that measurement finds the system in the state $\ket{v_i}$~\cite{b2}.

\subsection{Quantum Operators and Unitarity}\label{sec:Quantum Operators and Unitarity}
Quantum operators are linear transformations that act on states. An operator $\hat{A}$ maps $\ket{\psi}$ to another state $\hat{A}\ket{\psi} = \ket{\phi}$. For quantum computation, the operators must be unitary, satisfying $U^\dagger U = I$,
ensuring:
\begin{itemize}
    \item Reversibility: Reversible operations preserve all quantum information throughout the computation, which is vital not only for algorithmic correctness but also for maintaining quantum coherence. Moreover, irreversible operations would imply loss of information, which contradicts the deterministic evolution of isolated quantum systems.
    \item Probability conservation: Unitary operators preserve the inner product structure of Hilbert space, which ensures that the norm of the state vector remains constant during evolution~\cite{b2}.
\end{itemize}

\subsection{Key Quantum Gates}\label{sec:Key Quantum Gates}
Quantum gates manipulate qubit states analogously to classical logic gates but with additional capabilities (e.g., phase shifts, superposition)~\cite{b2}. Two critical single-qubit gates are:
Hadamard gate (H) which creates superposition from computational basis states as shown in Equation~\ref{eq:hadamart}:
    
\begin{equation}\label{eq:hadamart}
    H = \frac{1}{\sqrt{2}} \left( \begin{matrix}
    1&1  \\ 
    1&-1 \\ 
   \end{matrix}\right)
\end{equation}

\begin{tabbing} 
    where \hspace{0.6cm} \= $H\ket{0} = \ket{+}$\\
    \> $H\ket{1} = \ket{-}$\\
    \> $\ket{\pm} = \frac{\ket{0}\pm\ket{1}}{\sqrt{2}}$\\
\end{tabbing}
    
Phase gate (P) which introduces a relative phase $\theta$ to $\ket{1}$ as shown in Equation~\ref{eq:phase}:
    
\begin{equation}\label{eq:phase}
    P=\left( \begin{matrix}
    1&0  \\ 
    0&e^{i\theta} \\ 
   \end{matrix}\right)
\end{equation}

\begin{tabbing} 
    where \hspace{0.6cm} \= $P\ket{0} = \ket{0}$\\
    \> $P\ket{1} = e^{i\theta}\ket{1}$\\
\end{tabbing}

\section{Range Doppler Algorithm}\label{sec:Range Doppler Algorithm}
In SAR imaging systems, microwave pulses are transmitted from an airborne or spaceborne platform towards the target area. The backscattered echoes are collected and sampled, producing a two-dimensional raw signal $s(\tau, \eta)$ where $\tau$ represents the range (fast-time) dimension and $\eta$ denotes the azimuth (slow-time) dimension.
The \textbf{Range Doppler Algorithm (RDA)} processes these data into a focused image through sequential range and azimuth compression, leveraging Fourier-domain transformations. The algorithm achieves this through the efficient utilization of Fast Fourier Transforms (FFTs).

\subsection{Classical RDA}\label{sec:Classical RDA}
The classical RDA consists of four stages~\cite{b4}~\cite{b5} as depicted in Figure~\ref{fig:RDA}:
\begin{enumerate}
\item \textbf{Range Compression}: is the first major processing step in the Range-Doppler Algorithm (RDA), and its primary goal is to enhance range resolution by concentrating the dispersed pulse energy reflected from each target into a narrow peak. This step significantly improves the signal-to-noise ratio (SNR) which quantifies how much a signal stands out from background noise,  and prepares the data for further processing such as Range Cell Migration Correction (RCMC) and azimuth compression.

In most practical SAR systems, this operation is performed via frequency-domain matched filtering. Instead of correlating the received signal with the time, reversed transmitted pulse directly in the time domain, a computationally intensive process—this filtering is efficiently implemented in the frequency domain using the convolution theorem. The signal is first transformed into the frequency domain using the FFT, then multiplied by a range reference function $G(f_\tau)$, and finally transformed back using the inverse FFT:

\begin{equation}\label{eq:rangecompress}
s_{rc}(\tau, \eta) = \text{IFFT}_\tau \left[ \text{FFT}_\tau [s(\tau, \eta)] \cdot G(f_\tau) \right]
\end{equation}

where $G(f_\tau)$ is the range reference function, $(\tau, \eta)$ is the received raw signal, $\tau$ is the fast time (range), $\eta$ is the slow time (azimuth). This operation collapses all targets with identical slant ranges into single trajectories while preserving phase information.

The matched filter maximizes the SNR by aligning the received signal with the known transmitted pulse in phase, effectively compressing the wide-bandwidth chirp signal into a sharp spike in time. As a result, echoes from targets at the same slant range but different azimuth positions become concentrated along single trajectories in the 2D signal space, while maintaining phase information crucial for later steps such as interferometry or Doppler processing.

From a signal processing perspective, this filtering suppresses range ambiguities and background noise, making it easier to detect true target reflections. The output signal after range compression reveals the relative distances to the targets by the locations of the resulting peaks in the time domain. The peak amplitudes reflect the intensity of the backscattered energy, which is influenced by both the target's radar cross-section and its distance from the radar.

Visually, the range-compressed data appears as a superposition of hyperbolic arcs in the $(\tau, \eta)$ plane, where each arc corresponds to a ground target. These arcs represent the geometric locus of reflections from a target as the radar platform moves through the synthetic aperture.

\begin{figure}[ht!]
\begin{center}
		\includegraphics[width=1.0\columnwidth]{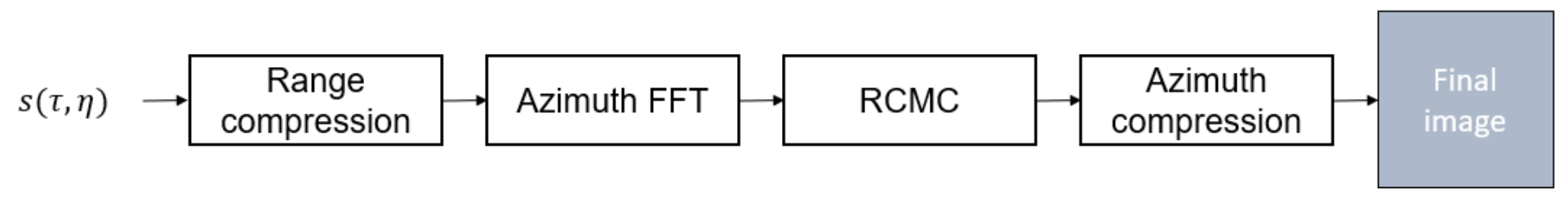}
	\caption{Block diagram of the classical Range Doppler Algorithm (RDA) used in Synthetic Aperture Radar (SAR) imaging. }
\label{fig:RDA}
\end{center}
\end{figure}

\item \textbf{Azimuth FFT}: The range-compressed signal is transformed to the azimuth frequency domain as expressed in Equation~\ref{eq:azimuthfft}:

\begin{equation}\label{eq:azimuthfft}
s_1(\tau, f_\eta) = \text{FFT}_\eta [s_{rc}(\tau, \eta)]
\end{equation}

\item \textbf{Range Cell Migration Correction (RCMC)}: is a critical step in the \textbf{Range Doppler Algorithm (RDA)}, addressing the range migration effect caused by the relative motion between the radar platform and the ground targets during the synthetic aperture. Due to this relative motion, echoes from a single point target do not remain in a fixed range bin over the aperture time but instead migrate across multiple bins. If uncorrected, this effect degrades image focus and resolution.

RCMC realigns these dispersed echoes by applying a phase correction in the range frequency domain, effectively remapping the signal so that each target's energy is concentrated in a single range cell. This correction allows subsequent azimuth compression to be applied correctly, ensuring that the resulting image is sharp and geometrically accurate.

Mathematically, the correction is applied as a linear phase shift in the range frequency domain:~\ref{eq:rcmc}

\begin{equation}\label{eq:rcmc}
    G_{\text{RCMC}}(f_r) = \exp\left[4i\pi \frac{f_r}{c}\left(R_0(\frac{1}{D(f_\eta,V)}-1\right)\right]
\end{equation}

where $D(f_\eta,V) = \sqrt{1-(\lambda f_\eta/2V)^2}$ is the Doppler compression factor, $V$ is the platform velocity, and $R_0$ is the reference range, $f_r$ is the range frequency, and $f_\eta$ is the azimuth (Doppler) frequency,.
This exponential term serves as a frequency-dependent phase multiplier that corrects for the slant-range curvature introduced by platform motion. The Doppler compression factor $D(f_\eta, V)$ accounts for the nonlinearity in the trajectory of target returns. As a result, RCMC flattens the curved migration path of target echoes, allowing them to be coherently integrated in the azimuth direction.

In implementation, this correction is often performed in the frequency domain using FFTs, as it allows for efficient convolution. The phase shift translates into a range shift in the time domain, aligning all backscattered signals from a single target.

An intuitive way to understand RCMC is to imagine the radar moving past a stationary object. At the start of the aperture, the object appears in one range bin, but as the radar moves, the geometry changes and the object's echo appears to "drift" through other range bins. RCMC "undoes" this drift, essentially straightening a curved path into a linear one~\cite{b6}.

\item \textbf{Azimuth Compression}:is the final step in the Range-Doppler Algorithm (RDA), responsible for focusing the radar returns in the azimuth (along-track) direction and producing the final high-resolution SAR image. Just as range compression improves resolution in the range dimension by applying matched filtering, azimuth compression performs an analogous operation in the Doppler (frequency) domain of azimuth.

After the Range Cell Migration Correction (RCMC) step, the radar data is organized in the range-Doppler domain. Each row of this two-dimensional dataset corresponds to a specific slant range, and each column represents a Doppler frequency (related to azimuth). The azimuth signal for each range bin must now be focused to concentrate the target energy into a sharp point in azimuth.

This is achieved through matched filtering in the azimuth frequency domain. The process applies a complex exponential filter designed to reverse the phase evolution imparted by the relative motion between the radar and the target. The phase history $\phi(f_a)$ of a stationary target during the synthetic aperture is typically quadratic, and correcting it allows for coherent integration of the received echoes. The operation is expressed mathematically as:~\ref{eq:azimuthcompress}:

\begin{equation}
s_{ac}(\tau, \eta) = \text{IFFT}_\eta \left[ s_2(\tau, f_\eta) \cdot H(f_\eta) \right]
\label{eq:azimuthcompress}
\end{equation}

$s_2(\tau, f_\eta)$ is the signal after range compression and RCMC, $H(f_\eta) = e^{-j\phi(f_\eta)}$ is the azimuth matched filter, $\eta$ is azimuth (slow) time, $f_\eta$ is the azimuth (Doppler) frequency.

The matched filter is given by: $H(f_a) = e^{-j\phi(f_a)}$, where $\phi(f_a)$ is the phase history of the target in the azimuth frequency domain
The matched filter $H(f_\eta)$ cancels out the phase modulation imposed by the radar-target geometry, which is typically parabolic due to the curved relative motion of the radar platform. By applying this filter, the algorithm aligns the phases of all echo contributions from the same target, thereby concentrating their energy and enhancing image sharpness in azimuth.

Following the application of the filter, an inverse FFT (IFFT) is used to transform the frequency-domain signal back to the azimuth time domain. This step completes the formation of the SAR image, delivering a focused two-dimensional representation of the scene with high resolution in both range and azimuth.

The effectiveness of azimuth compression is crucial for fine resolution and image clarity. Without it, the energy of each target would be spread out along the azimuth dimension, resulting in blurred and unfocused images. This step not only ensures geometric accuracy but also plays a pivotal role in improving contrast and enabling applications such as change detection or interferometric SAR (InSAR)~\cite{b7}.
\end{enumerate}

\subsection{Quantum RDA}\label{sec:Quantum RDA}
\begin{figure}[ht!]
\begin{center}
		\includegraphics[width=1.0\columnwidth]{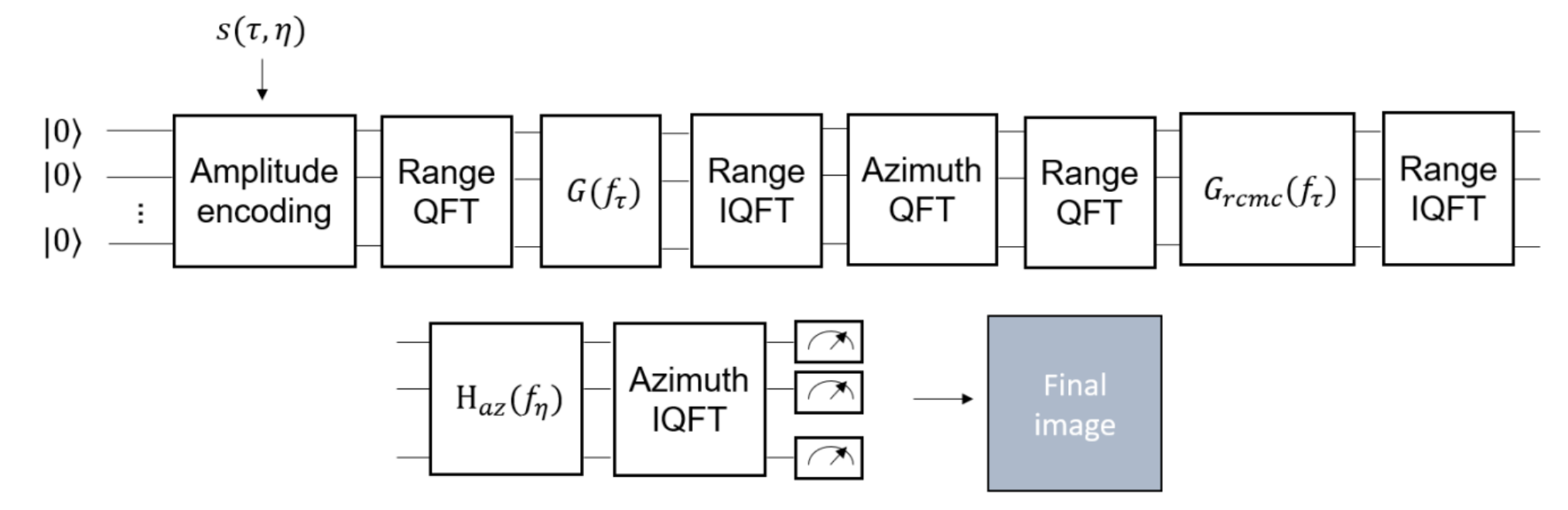}
	\caption{Quantum Range-Doppler Algorithm (QRDA): Proposed quantum circuit implementation of the classical Range-Doppler Algorithm (RDA).}
\label{fig:QRDA}
\end{center}
\end{figure}

Building upon the classical RDA framework, the proposed quantum implementation replaces FFT operations with the  \textbf{Quantum Fourier Transform (QFT)}, offering an exponential speedup in theory. This also means that the whole algorithm from encoding to measurement must be performed in the quantum domain to actually achieve this speedup~\cite{b1}. The proposed approach is depicted in Figure~\ref{fig:QRDA} where the classical information is first encoded into the quantum domain using amplitude encoding. The next steps involve applying the range compression filter, the RCMC filter, and the azimuth compression filter after moving to the frequency domain and back using QFT (and inverse QFT) applied to the range and azimuth lines, respectively. More details are provided in the following.

\subsection{Amplitude Encoding}\label{sec:Amplitude Encoding}
The first step in our algorithm is the encoding step, where we relied on amplitude encoding technique. Amplitude encoding embeds the information into the probability amplitudes of quantum states, which is done by first normalizing the dataset and then initializing the values into amplitudes. In our case, the encoded information consists of the SAR samples $s(\tau, f_\eta)$, made up of complex numbers that contain norm and phase information for each range and azimuth sample. Amplitude encoding is especially powerful because it allows one to encode N features with only $\log_2(N)$ qubits; for example, a 16x16 image would require only 16 qubits. 
Given a classical vector $\mathbf{x}=(x_0, x_1, ..., x_{N-1})$ representing our data, we encode it in a normalized quantum state as shown in Equation~\ref{eq:datarepresentation}:

\begin{equation}
\ket{\psi} = \sum_{i=0}^{N-1} x_i \ket{i}
\label{eq:datarepresentation}
\end{equation}

where $\ket{i}$ are the computational basis states of an n-qubit system (for $N= 2^n$).
The coefficients $x_i$  are the amplitudes that encode the classical data.
The vector must be normalized, that is, $\sum_{i} |x_i|^2 = 1$~\cite{b8}~\cite{b9}~\cite{b10}.

\section{QFT and QRCMC}\label{sec:QFT and QRCMC}

After successfully encoding the SAR raw data into quantum states, we can proceed with the application of the Quantum Fourier Transform (QFT) and the Quantum Range Cell Migration Correction (QRCMC). The QFT allows efficient transformation into the frequency domain, while the QRCMC applies precise phase corrections to compensate for range cell migration due to target motion.

\subsection{Quantum Fourier Transform}\label{sec:Quantum Fourier Transform}

The  \textbf{Quantum Fourier Transform (QFT)} is the quantum counterpart of the classical Discrete Fourier Transform (DFT). For $N=2^n$ samples, the \textbf{QFT} achieves an exponential speedup over the Fast Fourier Transform (FFT), reducing complexity from $O(n\log n)$ to $O(\log^2n)$ by exploiting quantum superposition and entanglement.
The \textbf{QFT} transforms a computational basis state $\ket{x}$ (where $x \in \{ 
0,1,\dots,N-1 \}$) into a superposition of Fourier basis states as depicted in Equation~\ref{eq:qft}:

\begin{equation}\label{eq:qft}
    \text{QFT}\ket{x} = \frac{1}{\sqrt{N}}\sum^{N-1}_{k=0}e^{2\pi ixk/N}\ket{k} 
\end{equation}

where $k$ represents the frequency components for an $n$-qubit system~\cite{b11}~\cite{b12}.

\begin{figure}[ht!]
\begin{center}
		\includegraphics[width=1.0\columnwidth]{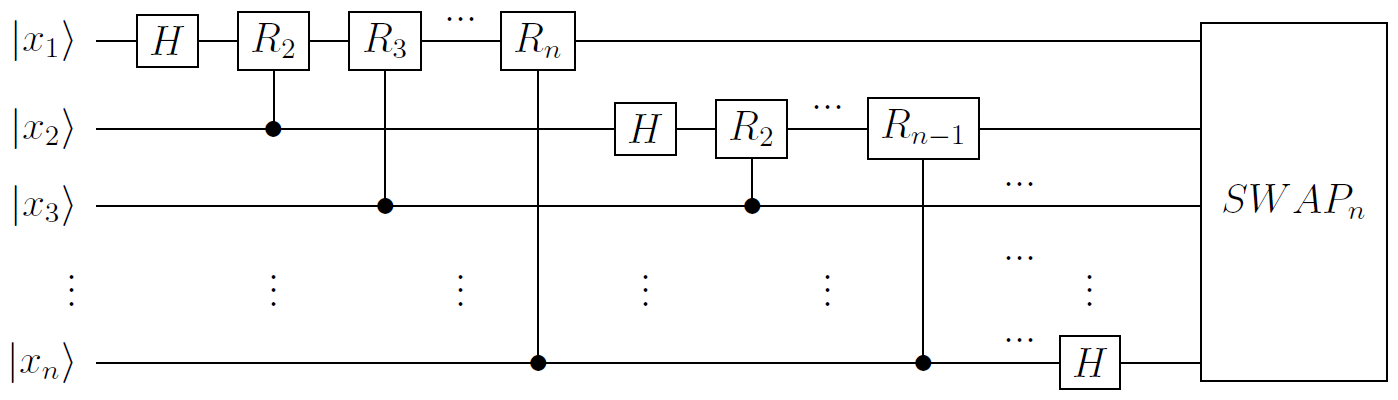}
	\caption{Quantum Fourier Transform Gate.}
\label{fig:QFT}
\end{center}
\end{figure}

The \textbf{QFT circuit} is constructed recursively using:
\begin{enumerate}
    \item Hadamard gates ($H$) previously discussed in Equation~\ref{eq:hadamart}.
    \item Controlled phase rotations to encode frequency-dependent phases defined in Equation~\ref{eq:phaserotation}:

\begin{equation}
    R_k=\left( \begin{matrix}
    1&0\\
    0&e^{2\pi i /2^k}
\end{matrix} \right) 
\label{eq:phaserotation}
\end{equation}

\end{enumerate}
The structure of the circuit reflects a recursive decomposition of the Fourier transform, with each qubit undergoing a Hadamard gate followed by progressively finer phase rotations conditioned on higher-order qubits, as illustrated in Figure~\ref{fig:QFT}~\cite{b13}.

\subsection{Quantum RCMC}\label{sec:Quantum RCMC}
Seeing that the \textbf{RCMC} filter is a correctional phase shift that realigns received signals shifted due to a moving target or radar, our aim will be to create a quantum gate that implements the phase shift coefficients that we have calculated classically. The \textbf{RCMC} filter, previously discussed in Equation~\ref{eq:rcmc}, is an array of phase elements that will filter the main radar data once multiplied, where each \textbf{RCMC} element needs to be multiplied by every range line.

To implement this logic in our quantum circuit, we created a gate that applies the corresponding phase shift to each of the amplitude encoded data, respectively. Being a phase-only array, implementing the filter into a gate that acts on the quantum states will not alter any of the probability amplitudes. In fact, making a diagonal matrix out of the \textbf{RCMC} elements (duplicated because each element corresponds to a range line and not a single sample) will give a reversible unitary gate, which is exactly what we need. Each element on the diagonal will be multiplied by the phase of the corresponding state. \textbf{RCMC} implementation is hence proposed here as a diagonal unitary operation as introduced in Equation~\ref{eq:rcmcunitary}:

\begin{equation}
U_{\text{RCMC}} = \bigoplus_{k=1}^{N_r} e^{i\Theta_k} \otimes I_{N_a}
\label{eq:rcmcunitary}
\end{equation}

where $\Theta_k$ contains the phase corrections for the $k$
-th range bin, and $I_{N_a}$ is the identity on azimuth qubits.

\begin{figure}[ht!]
\begin{center}
		\includegraphics[width=1.0\columnwidth]{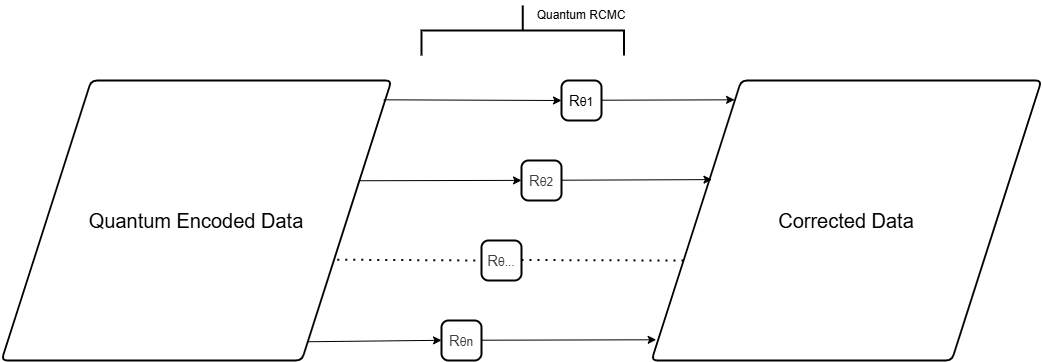}
	\caption{Schematic implementation of the quantum RCMC filter.}
\label{fig:qrcmc}
\end{center}
\end{figure}

A schematic implementation of this operation is shown in Figure~\ref{fig:qrcmc} where the quantum-encoded data are routed through a bank of rotation gates $R_k$, each applying the phase factor $e^{i\Theta_k}$ associated with a given range bin. The outputs of these rotations constitute the RCMC-corrected quantum state that is then used in the quantum RDA processing.

\subsection{Circuit Realization}\label{sec:Circuit Realization}
For a minimal 2×2 example (2 range bins × 2 azimuth samples) illustrated in Figure~\ref{fig:RCMC}, the \textbf{RCMC} operator takes the following form shown in Equation~\ref{eq:urcmc}:

\begin{equation}
\label{eq:urcmc}
U_{\text{RCMC}} =
\begin{pmatrix}
e^{i\theta_1} & 0 & 0 & 0 \\
0 & e^{i\theta_1} & 0 & 0 \\
0 & 0 & e^{i\theta_2} & 0 \\
0 & 0 & 0 & e^{i\theta_2}
\end{pmatrix}
\end{equation}

When applied to a state $\ket{\psi} = \begin{bmatrix}
\alpha_1 , \alpha_2 , \alpha_3 , \alpha_4 
\end{bmatrix}^T$, it yields to results as in Equation~\ref{eq:urcmcstate}:

\begin{equation}\label{eq:urcmcstate}
U_{\text{RCMC}}\ket{\psi} = \begin{bmatrix}
e^{\theta_1} \alpha_1\\
e^{\theta_1} \alpha_2\\
e^{\theta_2} \alpha_3\\
e^{\theta_2} \alpha_4
\end{bmatrix}
\end{equation}

Where $\alpha_i$ is the probability amplitude of each state $|i\rangle$, which contains the radar phase data, and $e^{\alpha_k}$ is the \textbf{RCMC} filter element acting on each range line k (containing 2 samples in this case).

\section{Results}\label{sec:Results}

We validated our quantum \textbf{RCMC} (Range Cell Migration Correction) implementation using the Qiskit \texttt{AerSimulator}, a state vector simulator that emulates an ideal, noise-free quantum environment~\cite{b14}. 

\begin{figure}[ht!]
\begin{center}
		\includegraphics[width=1.0\columnwidth]{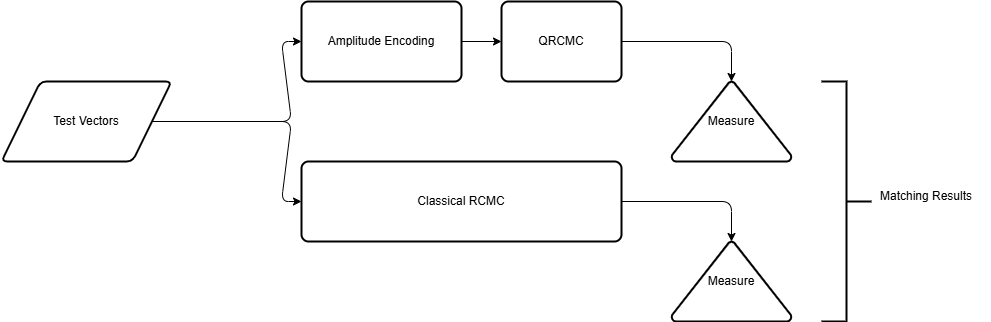}
	\caption{Isolation test used to validate the quantum \textbf{RCMC} implementation.}
\label{fig:isolation_diagram}
\end{center}
\end{figure}

To assess correctness, we constructed an \textbf{isolation test} in which the same complex-valued test vector, derived from a 64$\times$64 pixel subset of Sentinel-1 SAR data is processed in parallel by both the classical and quantum pipelines~\cite{b15}. As illustrated in Figure~\ref{fig:isolation_diagram}, the test vector is first amplitude-encoded, then passed through the proposed quantum \textbf{RCMC} gate, and finally measured. In the classical branch, the conventional \textbf{RCMC} operation is applied directly to the same input vector. The two outputs are then compared sample-wise in phase.

As shown in Figure~\ref{fig:results}, the phase difference plot reveals near-perfect alignment between the classical and quantum outputs. This confirms that our quantum algorithm reproduces the expected classical behavior. The minor discrepancies observed can be attributed to numerical precision limitations inherent in floating-point calculations, rather than errors in algorithmic logic. This test demonstrates the functional correctness of the proposed quantum \textbf{RCMC} algorithm \textbf{in isolation}.

To evaluate the quantum \textbf{RCMC}'s real-world applicability, we integrated the QRCMC gate into a full classical Range Doppler Algorithm (RDA) processing chain. For this experiment, we selected a smaller 8$\times$8 subset of real Sentinel-1 RAW data. This size was chosen deliberately to remain within the practical limits of current quantum simulators, which are constrained by exponential growth in memory and computation time as the number of qubits increases. Despite the reduced dimensionality, this test case is sufficient to validate the algorithm's end-to-end correctness.

\begin{figure}[ht!]
\begin{center}
		\includegraphics[width=1.0\columnwidth]{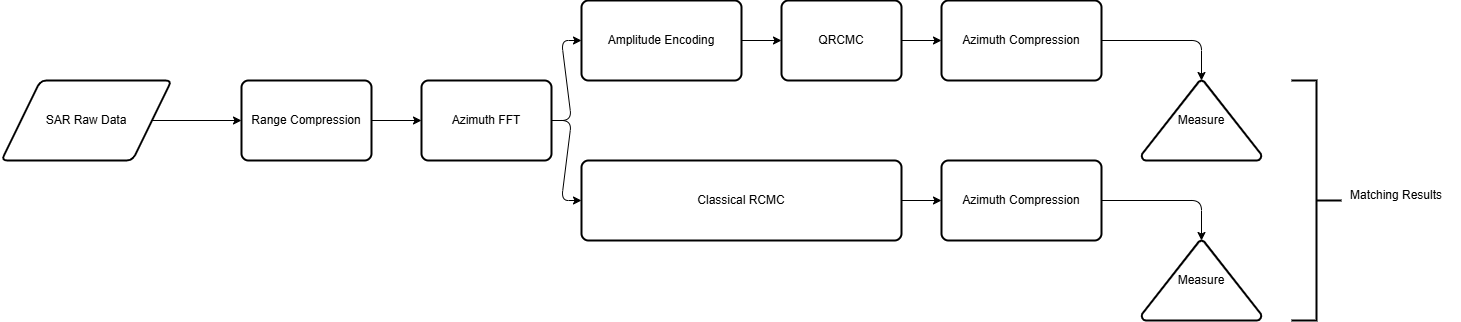}
	\caption{Integration of the quantum \textbf{RCMC} into a full RDA processing chain.}
\label{fig:integration_diagram}
\end{center}
\end{figure}

The hybrid processing flow is depicted in Figure~\ref{fig:integration_diagram}. Starting from the SAR RAW data, both branches share the initial classical stages: range compression followed by an azimuth FFT. At this point, the pipeline splits into two paths. In the \emph{quantum} branch, the intermediate data are amplitude-encoded, processed by the QRCMC gate, and then passed through classical azimuth compression before measurement. In the \emph{classical} branch, the conventional RCMC is applied directly, followed by the same azimuth compression and measurement. The two measured outputs are then compared on a pixel-by-pixel basis.

As illustrated in Figure~\ref{fig:iresults}, the resulting phase difference again demonstrates excellent alignment between the quantum-integrated and purely classical outputs. Similar to the isolated test case, any observed discrepancies are minor and attributable to floating-point numerical precision, not algorithmic faults. These results confirm that the quantum \textbf{RCMC} component can be correctly embedded within a full RDA workflow and operate in conjunction with classical stages, thereby demonstrating its functional reliability \textbf{under realistic conditions}.

\begin{figure}[ht!]
\begin{center}
		\includegraphics[width=1.0\columnwidth]{figures/results.png}
	\caption{Phase difference between classical and quantum RDA outputs for a 64×64 Sentinel-1 subset. The close match confirms the quantum implementation's accuracy.}
\label{fig:results}
\end{center}
\end{figure}

\begin{figure}[ht!]
\begin{center}
		\includegraphics[width=1.0\columnwidth]{figures/integration_results.png}
	\caption{Phase difference between classical and quantum-integrated RDA outputs for an 8×8 Sentinel-1 subset. The high degree of alignment confirms successful integration of the QRCMC module.}
\label{fig:iresults}
\end{center}
\end{figure}

\section{Conclusion}\label{sec:Conclusion}
This work demonstrates the feasibility of quantum acceleration for \textbf{Earth Observation} applications. Our implementation maintains mathematical equivalence with classical RDA while being fully executable on quantum hardware, as demonstrated through simulations of a 64×64 Sentinel-1 SAR subset.

Future work will focus on developing a complete, practical quantum RDA that integrates all processing steps within the quantum domain, as achieving a true speedup requires the successful implementation of the entire pipeline. Furthermore, future advancements in quantum sensors for satellites could eliminate the need for data encoding, further enhancing speed and efficiency.

However, it is essential to recognize that current progress is bounded by the limitations of Noisy Intermediate-Scale Quantum (NISQ) hardware. Our results were obtained through idealized simulations, which do not yet account for decoherence, gate errors, and readout noise that characterize present-day quantum devices. Addressing these challenges will be crucial for transitioning from proof-of-concept to deployable quantum SAR systems.

In conclusion, while we remain in the early stages of quantum SAR processing, our work offers a solid foundation and a roadmap toward scalable quantum solutions for remote sensing. As quantum hardware matures and algorithmic techniques evolve, the integration of quantum computing into Earth Observation may unlock new levels of performance, precision, and capability across a range of scientific and operational applications.

\newpage
\newpage

{
	\begin{spacing}{1.17}
		\normalsize
		\bibliography{authors} 
	\end{spacing}
}

\end{document}